\documentstyle[multicol,aps,prl,epsf]{revtex}
\begin{document}

\draft

\title{
$d_{x^2-y^2}$- vs. $d_{xy}$- like pairings 
in organic superconductors $\kappa$-(BEDT-TTF)$_2$X
}

\author{
Kazuhiko Kuroki$^1$, Takashi Kimura$^2$, Ryotaro Arita$^3$, 
Yukio Tanaka$^4$, and Yuji Matsuda$^5$
}

\address{
$^1$ Department of Applied Physics and Chemistry, the University of
Electro-communications, Chofugaoka, Chofu-shi, Tokyo 182-8585, Japan\\
$^2$ NTT Basic Research Laboratories, NTT Corporation, Morinosato-Wakamiya,
Atsugi 243-0198, Japan\\
$^3$ Department of Physics, University of Tokyo, Hongo, Tokyo 113-0033, Japan\\
$^4$ Department of Applied Physics, Nagoya University, Nagoya 464-8063, Japan\\
$^5$ Institute for Solid State Physics, University of Tokyo, Kashiwanoha,
Kashiwa, Chiba 277-8581, Japan.
}

\date{\today}

\maketitle

\begin{abstract}
Pairing symmetry in $\kappa$-(BEDT-TTF)$_2$X 
is re-examined by applying fluctuation exchange method 
to the Hubbard model on the original four band lattice for the first time.
Our study is motivated by the discrepancy between 
a recent thermal conductivity/tunneling spectroscopy 
experiments and the previous theories based on the effective 
single or two band model. In contrast to the previous theories, 
we find that $d_{x^2-y^2}$-like pairing dominates 
over $d_{xy}$ in agreement with 
recent experiments. Our study provides a general lesson that 
even if the Fermi surface seems to be reproduced by an effective model, 
the pairing symmetry can be different from that of the original system.
\end{abstract}

\medskip

\pacs{PACS numbers: 74.20.Rp, 74.70.Kn, 74.20.Mn}

\begin{multicols}{2}
\narrowtext
Unconventional mechanisms of superconductivity is one of the most
challenging issues in solid state physics. Not only that there is 
a possibility of accomplishing high $T_c$ as in the cuprates,
but also the variety of pairing symmetries provide rich physics.

Recently, it has become increasingly clear that organic materials
can provide various stages for unconventional pairings.
In particular, $\kappa$-(BEDT-TTF)$_2$X is of special interest 
in that superconductivity lies near the antiferromagnetic
insulating phase like in the cuprates,\cite{Review} 
suggesting that spin fluctuations 
are playing important role in the occurrence of superconductivity.

In order to clarify the mechanism of superconductivity,
it is crucial to identify the pairing symmetry.
Theoretically, previous studies on the Hubbard model with 
fluctuation exchange (FLEX) approximation\cite{Schmalian,KK,KM} or 
quantum Monte Carlo calculation\cite{KA} have suggested 
that superconductivity in $\kappa$-(BEDT-TTF)$_2$X is 
similar to the $d$-wave superconductivity in the cuprates.
However, a recent thermal conductivity\cite{IM} as well as 
tunneling spectroscopy measurements\cite{Arai} casts doubt on such a 
view, which is exactly the motivation of the present study.

\begin{figure}
\begin{center}
\leavevmode\epsfysize=50mm \epsfbox{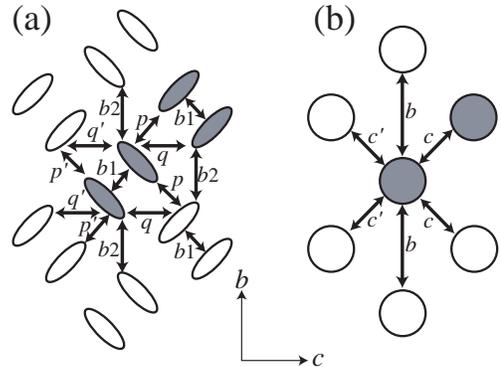}
\caption{The lattice structure of $\kappa$-(BEDT-TTF)$_2$X.
(a) the original four band lattice, and (b) the effective 
two band lattice. $b1, b2,\cdots$ stand for the hopping integrals 
$t_{b1}, t_{b2},\cdots$. A unit cell consists of the hatched molecules 
(or sites).
}
\label{fig1}
\end{center}
\end{figure}

\begin{figure}
\begin{center}
\leavevmode\epsfysize=50mm \epsfbox{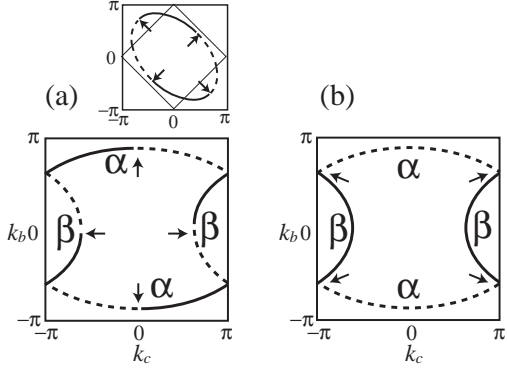}
\caption{(a) $d_{xy}$-like and (b) $d_{x^2-y^2}$-like 
gap functions on the Fermi surfaces. 
Solid (dashed) lines represent positive (negative) values.
The inset in (a) shows the Fermi surface and the 
$d_{xy}$-like gap function in the unfolded Brillouin zone,
which can be obtained by moving the right and the left portions
of the $\beta$ band Fermi surface in the folded 
Brillouin zone by $(-2\pi,0)$ and $(+2\pi,0)$, respectively, 
and rotating it by 45 degrees.
Arrows show the position of the nodes of the gap function.
Note that the points where the $\alpha$ and the $\beta$ 
band Fermi surfaces (almost) touch in (a) are not the nodes. 
There, the gap function abruptly jumps from a negative to a 
positive value (or vise versa).
This can be understood by looking at the corresponding points 
in the unfolded Brillouin zone.
}
\label{fig2}
\end{center}
\end{figure}

Let us first briefly summarize the background of the issue.
As shown in Fig.\ref{fig1}(a), $\kappa$-(BEDT-TTF)$_2$X is essentially a 
four band system in the sense that there are four BEDT-TTF 
molecules per unit cell.
The lattice structure is dimerized in that the hopping integral in the 
$b$-direction alternates as $t_{b1}, t_{b2}, t_{b1},\cdots$, 
where $|t_{b1}|>|t_{b2}|$. The band filling is $n=1.5$, where $n$ is 
defined as $n=$(the number of electrons/ the number of molecules).
The upper two bands, which we will refer to as $\alpha$
and $\beta$ bands hereafter, cross the Fermi level, so 
the Fermi surface consists of two portions (see Fig.\ref{fig2}).
In the limit of large $t_{b1}$, namely, when the dimerization is strong, 
each dimer can be considered as a single lattice site, so 
the system reduces to a two band model shown in Fig.\ref{fig1}(b)
with $n=1$, where the band filling is now defined as 
$n$=(the number of electrons/the number of {\it sites}).\cite{Tamura,KF} 
In other words, the energy gap between the upper two and the lower
two bands becomes large when the dimerization is strong, so that the 
lower two bands, which do not cross the Fermi level, can be neglected. 
In this limit, the effective hopping integrals $t_b$ and $t_c$ are given 
as $t_b=-t_{b2}/2$, $t_c=(-t_p+t_q)/2$, and $t_{c'}=(-t_{p'}+t_{q'})/2$
in terms of the original hopping integrals,\cite{Tamura} which gives 
$|t_b/t_c|\sim 0.8$.
The system further reduces to a single band model when $t_c=t_{c'}$. 

Assuming that the dimerization is sufficiently strong in 
$\kappa$-(BEDT-TTF)$_2$X, 
the two band\cite{Schmalian} or the single band Hubbard model\cite{KK,KM,KA} 
has been adopted in the previous theoretical studies. 
There, it has turned out that antiferromagnetic spin fluctuations 
lead to an anisotropic superconductivity with a gap function 
of the form shown
in Fig.\ref{fig2}(a), which we will refer to as 
$d_{xy}$-like pairing hereafter from the direction of the nodes.
This gap function
\noindent
is similar to that of the $d$-wave 
superconductivity in the high $T_c$ cuprates, as can be seen more clearly by 
unfolding the Brillouin zone (inset of Fig.\ref{fig2}(a)).

Experimentally, NMR experiments \cite{Mayaffre,Soto,Kanoda2}
have indeed supported the spin-fluctuation scenario,
but those experiments do not give direct information on the position of 
the nodes in the gap function. 
A millimeter-wave transmission experiment has suggested a 
$d_{xy}$-like gap function,\cite{Schrama} but 
different interpretations on this experiment have recently been 
proposed,\cite{Hill,Shibauchi} so the situation has not been settled.

Quite recently, tunneling spectroscopy\cite{Arai} 
and thermal conductivity measurements\cite{IM} have been 
performed for (BEDT-TTF)$_2$Cu(NCS)$_2$ in an attempt to 
determine the anisotropy in the gap function.
Surprisingly, both of the experiments have suggested that
the gap function is more like Fig.\ref{fig2}(b) 
(referred to as $d_{x^2-y^2}$-like hereafter) than (a),
casting doubt on the spin-fluctuation scenario based on the single band 
or the two band approximation.

In the present study, we apply FLEX to the Hubbard model,
$H=\sum_{i,j}\sum_{\sigma=\uparrow,\downarrow}t_{ij}
(c_{i\sigma}^\dagger c_{j\sigma}+{\rm H.c})
+U\sum_i n_{i\uparrow}n_{i\downarrow}$,
on the {\it original four band lattice} (Fig.\ref{fig1}(a)) 
for the first time.
Surprisingly enough, our results show that $d_{x^2-y^2}$-like 
pairing dominates $d_{xy}$
when dimerization is not so strong, as in $\kappa$-(BEDT-TTF)$_2$Cu(NCS)$_2$.
The present study provides a general lesson that 
even if the Fermi surface seems to be well reproduced by an effective 
model, its pairing symmetry can be quite different from that of the 
original system.
\begin{figure}
\begin{center}
\leavevmode\epsfysize=155mm \epsfbox{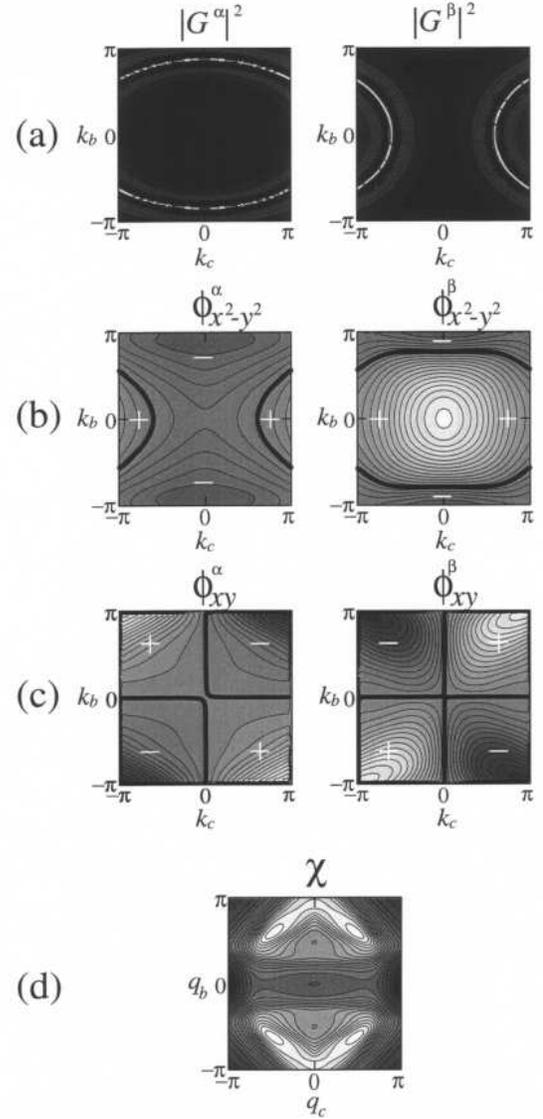}
\caption{Contour plots (for the lowest Matsubara frequency)
of (a)$|G^\nu|^2$, (b) $\phi^\nu_{x^2-y^2}$, 
(c) $\phi^\nu_{xy}$ $(\nu=\alpha,\beta)$ and (d)$\chi$ with 
$t_{b1}=-1.0$, $t_{b2}=-0.5$, $t_{p}=-0.45$, $t_{q}=0.15$,
$U=6$, and $T=0.01$. The thick lines in (b) and (c) represent 
the nodes of the gap function.}
\label{fig3}
\end{center}
\end{figure}

FLEX is a kind of self-consistent 
random phase approximation, which is known to be appropriate 
for treating strong spin fluctuations.\cite{Bickers,Grabowski,Dahm}
In the four band version of FLEX,\cite{Takashi} the Green's function $G$, 
the susceptibility $\chi$, the self-energy $\Sigma$, and 
the superconducting gap function $\phi$ all become $4\times 4$ matrices,
e.g., $G_{lm}({\bf k}, i\varepsilon_n)$, where $l,m$ specify
the four sites in a unit cell.
The site-indexed matrices for Green's function and the gap functions
can be converted into band(denoted as $\alpha,\beta,\cdots$)-indexed 
ones with a unitary transformation. 
As for the spin susceptibility, we diagonalize the 
$4\times 4$ matrix $\chi_{\rm sp}$ and 
concentrate on the largest eigenvalue, denoted as $\chi$.  
We take $64\times 64$ $k$-point meshes and 
up to 4096 Matsubara frequencies. 
\begin{figure}
\begin{center}
\leavevmode\epsfysize=50mm \epsfbox{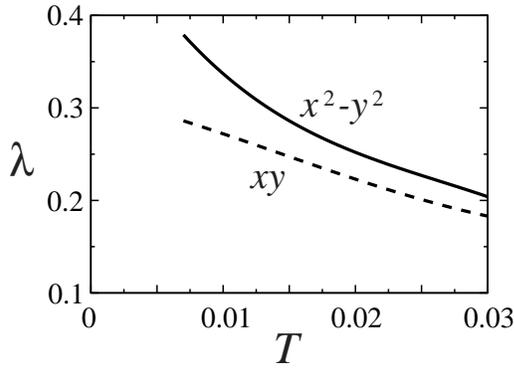}
\caption{$\lambda_{x^2-y^2}$ and $\lambda_{xy}$ plotted as functions 
of temperature for the same parameter values as in Fig.\ref{fig3}.}
\label{fig4}
\end{center}
\end{figure}

The superconducting gap function is obtained as the eigenfunction of the 
linearized Eliashberg equation. In order to solve the Eliashberg equation,
we use the power method, where 
$d_{x^2-y^2}$-like or $d_{xy}$-like gap functions has been assumed 
as a trial function. Consequently, we have obtained 
$d_{x^2-y^2}$ and $d_{xy}$-like gap functions $\phi_{x^2-y^2}^{\nu}$ 
and $\phi_{xy}^{\nu}$ $(\nu=\alpha,\beta,\cdots)$, respectively, 
together with the corresponding eigenvalues   
$\lambda_{x^2-y^2}$ and $\lambda_{xy}$.

We have modeled $\kappa$-(BEDT-TTF)$_2$Cu(NCS)$_2$ by taking 
$t_{b1}=-1$, $t_{b2}=-0.5$, $t_p=-0.45$, and $t_q=0.15$ in accordance with 
a band calculation result given in ref.\cite{Komatsu}.
(We have neglected the small difference between $t_q$ and $t_{q'}$ 
or between $t_p$ and $t_{p'}$.)
In Fig.\ref{fig3} (a), we show contour plots for $|G^\nu|^2$ 
$(\nu=\alpha,\beta)$, where the ridges represent the Fermi surfaces.
In (b) and (c), we show contour plots for $\phi_{x^2-y^2}^\nu$
and $\phi_{xy}^\nu$, respectively. Superposing the gap functions on the 
Fermi surfaces, it can be seen that $\phi_{x^2-y^2}$ and 
$\phi_{xy}$ essentially have the functional forms given in Fig.\ref{fig2}
(b) and (a), respectively.

The corresponding eigenvalues $\lambda_{x^2-y^2}$ 
and $\lambda_{xy}$ are plotted against temperature 
in Fig.\ref{fig4}. The tendency that 
$\lambda_{x^2-y^2}-\lambda_{xy}$ increases upon lowering the temperature
clearly demonstrates that $d_{x^2-y^2}$-like pairing dominates over
$d_{xy}$.

\begin{figure}
\begin{center}
\leavevmode\epsfysize=70mm \epsfbox{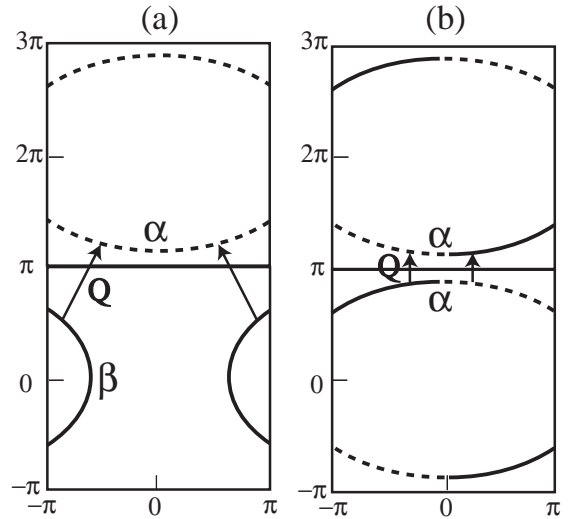}
\caption{(a) The nesting vector {\bf Q} bridges the $\alpha$ and 
the $\beta$ band portions of the Fermi surface when the 
dimerization is not so strong. 
(b)The nesting vector 
lies within the $\alpha$ band portion of the Fermi surface 
for the single band and the two band models.
The same applies for the four band model when the dimerization is strong. 
$d_{x^2-y^2}$- and $d_{xy}$-like gap functions are shown in (a)
and (b), respectively.
}
\label{fig5}
\end{center}
\end{figure}

The reason why $d_{x^2-y^2}$-like superconductivity dominates over $d_{xy}$ 
can be found in the spin structure.
Namely, the spin susceptibility $\chi$, shown in Fig.\ref{fig3}(d), 
peaks around ${\bf Q}\sim (\pm 0.5\pi,\pm 0.6\pi)$,\cite{comment2}
which is a consequence of partial nesting between the $\alpha$ 
and the $\beta$ band portions of the Fermi surface (Fig.\ref{fig5}(a)).
Thus, the interband pair scattering processes dominate over 
intraband ones, resulting in a gap function that has different signs between 
the two portions of the Fermi surfaces, but does not change 
sign within each portion.
\begin{figure}
\begin{center}
\leavevmode\epsfysize=35mm \epsfbox{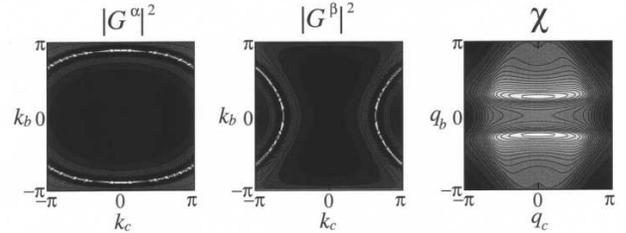}
\caption{Contour plots of $|G^\nu|^2$ and $\chi$ for the 
single band model with $t_b=0.83$, $t_c=1$, $U=6$ and $T=0.03$.
Folded Brillouin zone is adopted for clear comparison with 
Fig.\ref{fig3}
}
\label{fig6}
\end{center}
\end{figure}

The present FLEX results for the four band model is indeed different from 
those for the corresponding single band model. In Fig.\ref{fig6}, 
we show $|G|^2$ and $\chi$ for the single band Hubbard model 
with $n=1$, $t_b=0.83$, $t_c=t_{c'}=1$, and $U=6$, 
where the ratio $t_b/t_c$ is taken to be the same as 
$[-t_{b2}/2]/[(-t_p+t_q)/2]$ with $t_{b2}=-0.5$, $t_p=-0.45$ and $t_q=0.15$.
(We have adopted the folded Brillouin zone in order to make clear 
comparison with the results for the four band model.)
Although the Fermi surfaces look similar to that of the 
four band model (Fig.\ref{fig3}(a)), the spin susceptibility peaks 
around $(0,\pm 0.25\pi)$ in contrast to that of the four band model. 
Thus the pair scattering mainly occurs within the $\alpha$ band 
portion of the Fermi surface (Fig.\ref{fig5}(b)), and 
consequently $d_{xy}$-like superconductivity dominates 
in the single band model. 
A general lesson here 
is that even if the Fermi surfaces of an effective 
model `look similar' to that of the original system, 
\begin{figure}
\begin{center}
\leavevmode\epsfysize=42mm \epsfbox{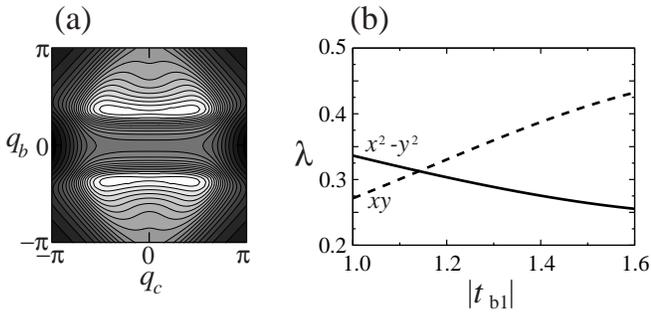}
\caption{(a)A contour plot of $\chi$ for 
the same parameter values as in Fig.\ref{fig3} except that 
$t_{b1}=-1.6$. (b)$\lambda_{x^2-y^2}$ and $\lambda_{xy}$ plotted 
against $|t_{b1}|$ at $T=0.01$.
}
\label{fig7}
\end{center}
\end{figure}
\noindent
the spin structure, 
and consequently the superconducting gap function, 
can be quite different.

When the dimerization is strong, the four band model should 
tend to the single band model.
In fact, if we take for example $t_{b1}=-1.6$ and leave the 
other parameters unchanged, $\chi$, shown in Fig.\ref{fig7}(a), 
now has a structure similar to the single band result Fig.\ref{fig6}(c). 
Consequently, $d_{x^2-y^2}$-like superconductivity gives way to $d_{xy}$
as seen in Fig.\ref{fig7}(b), 
where we have plotted $\lambda_{x^2-y^2}$ and $\lambda_{xy}$
against $|t_{b1}|$ at $T=0.01$.

As we have seen, the competition between $d_{x^2-y^2}$- and $d_{xy}$-like
pairings is sensitive to the parameter values. 
This means that the two types of superconductivity 
can possibly be nearly degenerate for certain combinations of the anion X,
\cite{comment} temperature, pressure, and/or magnetic field.
In such a case, $d_{x^2-y^2}+id_{xy}$ pairing, a pairing with 
broken time reversal symmetry, is highly probable 
since the gap is fully open in this channel.
Investigations on such a possibility would be interesting 
both theoretically and experimentally.

To summarize, we have examined the Hubbard model on the 
original four band lattice of $\kappa$-(BEDT-TTF)$_2$X.
We have found that $d_{x^2-y^2}$-like superconductivity 
dominates over $d_{xy}$ when the dimerization is not so strong,
as in $\kappa$-(BEDT-TTF)$_2$Cu(NCS)$_2$.
Our finding is consistent with recent thermal conductivity 
and tunneling spectroscopy measurements.
 
Discussions with Hideo Aoki are gratefully acknowledged.
Numerical calculations were performed at the Supercomputer Center,
ISSP, University of Tokyo, and at the Computer Center,
University of Tokyo. K.K. acknowledges 
a Grand-in-Aid for Scientific Research from the Ministry of 
Education.

\end{multicols}

\begin{references}
\bibitem{comment2} The spin structure of $\kappa$-(BEDT-TTF)$_2$X 
is known to be commensurate with the lattice 
in and near the insulating phase\cite{Review}. This is 
because the localized spin picture is valid in the insulating 
phase. However, when the system is metallic 
and the Fermi surface is well defined (as is expected near the 
superconducting phase), the spin structure should 
be affected by the nesting properties of the Fermi surface 
and in general can be incommensurate.
\end{references}
\end{document}